\documentclass[conference]{IEEEtran}
\usepackage{cite}
\usepackage[pdftex]{graphicx}
\usepackage{amsmath}
\usepackage{algorithmic}
\usepackage{array}
\usepackage{caption} 
\usepackage{amsmath}
\begin{document}
\IEEEoverridecommandlockouts

\title{
 Building test batteries based on analysing random number generator tests within the framework of algorithmic information theory}
\author{\IEEEauthorblockN{Boris Ryabko}
\IEEEauthorblockA{Federal Research Center for Information and Computational Technologies \\Novosibirsk, Russian Federation \\Email: boris@ryabko.net\\}
}

\maketitle

\begin{abstract}
The problem of testing random number generators is considered and it is shown that an approach based on algorithmic information theory allows us to compare the power of different tests in some cases where the available 
 methods of mathematical statistics do not distinguish between the tests.  In particular, it is shown that tests based on data compression methods using dictionaries should be included in the test batteries.
\end{abstract}

\IEEEpeerreviewmaketitle

\textbf{keywords:}   random number generator,  statistical test,  algorithmic information theory,  battery of tests, data compression, universal coding,    Hausdorff dimension.

\section{Introduction} 
Random numbers play an important role in cryptography, gambling, Monte Carlo methods and many other applications. Nowadays, random numbers are generated using so-called random number generators (RNGs), and the ``quality'' of the generated numbers is evaluated using special statistical tests \cite{l1}. 
This problem is so important for applications that there are special standards for RNGs and for so-called test batteries, that is,\ sets of tests. 
 \cite{l2007testu01, NIST}.
The current practice for using an RNG is to verify the sequences it generates with  tests from some battery (such as those recommended by   \cite{l2007testu01, NIST} or other standards).

Let us recall 
 what a statistical test is. In the problem under consideration, there is a main hypothesis $H_0 = \{$ the sequence $x$ is random $\}$ and an alternative hypothesis $H_1 = \neg H_0$. (In the probabilistic approach,  $H_0$ is that the sequence is generated by a Bernoulli source with equal probabilities 0 and 1.)
 %
 %
 %
  A test is an algorithm for which the input is the prefix $x_1 ... x_n$ (of the infinite  sequence $x_1,\dots,x_n,\dots $) 
  and the output is one of  two possible words: {\em random} or {\em non-random} (meaning that the sequence is random or non-random, respectively). 
Many statistical tests are designed to test some alternative hypotheses $H_1$ described as classes of random processes (e.g., Bernoulli process with unequal probabilities 0 and 1, Markov chains with some unknown parameters,  stationary ergodic processes, etc.) \cite{l1,l2007testu01, NIST,j4}.

A natural question is: how do we compare different tests and, in particular, create a suitable battery of tests? Currently, this question is mostly addressed experimentally: possible candidate tests are applied to a set of known RNGs and the  tests that reject more ("bad") RNGs are  suitable candidates for the battery. In addition, researchers try to choose independent tests (i.e., those that reject different RNGs) and take into account other natural properties (e.g., testing speed, etc.) \cite{l1,l2007testu01, NIST,j4}.
Obviously, such an approach depends significantly on the set of selected tests and RNGs pre-selected for consideration. It is worth noting that at present there are dozens of RNGs and tests, and their number is growing fast, so the recommended batteries of tests are rather unstable (see \cite{j4}).

The goal  of this paper is to develop a theoretical framework for test comparison and illustrate it by comparing some popular tests.  In particular, it is shown that dictionary-based statistical tests are useful for test batteries.
For this purpose we propose the following mathematical model.
We first generalise statistical tests to infinite sequences, and then use the Hausdorff dimension to compare the ``size'' of sets of  sequences accepted by the test for different tests. 
 This comparison allows us to rank the tests according to their performance. 

Based on the aprroach described, 
 we give  some practical recommendations for building test batteries.
 In particular, we recommend including in the test batteries a test based on a dictionary data compressor, like Lempel-Ziv codes \cite{ziv}, grammar-based codes \cite{gb2}  and some others. 

The rest of the paper consists is organized as follows.  The next part contains definitions and preliminary information, the third part is a comparison of the test performance on Markov processes with different memories and general stationary processes, and the fourth part investigates tests based on Lempel-Ziv data compressors. The fifth part is a brief conclusion; some of the concepts used in the paper are given in the Appendix.

\section{Definitions and preliminaries} 
\subsubsection*{Hypothesis testing}

Let there be a hypothesis $H_0$,    
 some alternative $H_1$,  
 let  $ T $ be a test and $\tau$ be a statistic, that is,  
a function on   $ \{0,1 \}^n $  which is applied to a binary sequence $ x = x_1 ... x_n $.
Here and below $\{0,1\}^n$ is the set of all $n$-bit binary words, $\{0,1\}^\infty$ is the set of all infinite  words 
$x_1x_2 ... , x_i \in \{0,1\}$.


By definition,  Type I error occurs   if
$H_0$ is true and  $H_0$ is  rejected;  the significance level is defined as the 
 probability  of the Type I error.
Denote the critical region of the test $T$ for the  significance level $\alpha$ by $\bar{C}_T(\alpha,n)$  and let 
$C_T(\alpha,n)$ $ =  \{0,1\}^n \setminus \bar{C}_T(\alpha,n) .$ 
  Recall, that, by definition,    
$H_0$ is rejected if and only if $x \in \bar{C}_T(\alpha,n)$ and, hence,
 \begin{equation}\label{alphasize}
|\bar{C}_T(\alpha,n)| \le 2^n \alpha \, ,
 \end{equation}
see \cite{ks}.
We also apply another natural limitation. We consider only tests $T$ such that 
 for all $n$ and $\alpha_1 < \alpha_2 $ $\,\,\, \bar{C}_T(\alpha_1, n) \subset \bar{C}_T(\alpha_2, n)$.
 (Here and below $|X|$ is the number of elements $X$ if $X$ is a set, and the length of $X$, if $X$ is a word.)  

 A finite sequence $x_1 ... x_n$ is considered  random for a given test $T$ and the significance level $\alpha$ if   
it belongs to $ C_T(\alpha,n)$. 

\subsection*{Batteries of tests.} 

Let us consider a situation where the randomness testing  is performed by conducting  a battery  of statistical  tests for randomness.  Suppose that the battery $\hat{T}$ contains 
a finite or countable set of 
 tests   $T_1, T_2, ... $ and  $\alpha_i$ is the significance level of $i-$th test, $i= 1,2,  ... $.  If the battery is applied in such a way that
 the hypothesis $H_0$ is rejected when at least one  test in the battery rejects it, then the significance level $\alpha$
of this battery satisfies the following inequality:
 \begin{equation}\label{a-g}
\alpha \le \sum_{i=1}^\infty \alpha_i \, , 
\end{equation}
because $P(A+B) \le P(A)+P(B) $ for any events $A$ and $B$. 

It will be convenient to formulate this inequality in a different way.
Suppose there  is  some $ \alpha  \in (0,1)$ and a sequence $ \omega$ of non-negative $ \omega_i$  such that $\sum_{i=1}^\infty
\omega_i \le 1$. For example, we can define the following sequence $\omega^*$:
\begin{equation}\label{ome}
\omega^*_i =1 / (i (i+1)) \, \,\,\, i=1,2, ... \,\, .
\end{equation}
If  the significance level $T_i$ equals $\alpha \omega_i$, then the significance level of the battery $\hat{T}$ is not grater than $\alpha$.
(Indeed, from (\ref{a-g}) we obtain $\sum_{i=1} \alpha_i  = \sum_{i=1} (\alpha \omega_i)$ $ =\alpha \sum_{i=1} \omega_i \le \alpha $.) 
Note that this simple observation makes it possible to treat a test battery as a single test. 

\subsubsection*{Random and non-random infinite sequences}
 Kolmogorov complexity is one of the central notations of algorithmic information theory (AIT), see \cite{hu,vi,ca,downey2006calibrating,merkle2006kolmogorov,vv,ve}. We will consider the so-called unprefixed Kolmogorov complexity $K(u)$, which is defined on finite binary words $u$ and is closely related to the notion of randomness.
   More precisely, an infinite binary sequence $x=x_1 x_2...$ is random if there exists a constant $C$ such that
\begin{equation}\label{kr}
n - K(x_1...x_n)   < C 
 \end{equation}
for all $n$, see \cite{enc}.
Conversely, the sequence $x$ is non-random if
$$ \forall \,\,	C >0\,\, \exists\, n_C  \,\,\, n_C - K(x_1...x_{n_C} )  \ge C   
$$
In some sense, Kolmogorov complexity is the length of the shortest lossless prefix-free code, that is, for any (algorithmically realisable) code $f$ there exists a constant $c_f$ for which
$K(u) \le |f(u)| +c_f $ \cite{hu,vi,ca,downey2006calibrating,merkle2006kolmogorov,vv,ve}.
Recall that a code $f$ is lossless if there is a mapping $f^{-1}$ such that for any word $u$ $\,\,f^{-1}(f(u)) = u$ and $f $ is unprefixed if for any words $u,v$, $f(u)$ is not a prefix of $f(v)$ and $f(v)$ is not a prefix of $f(u)$.  

Let $f$ be a lossless  unprefixed code  defined for all finite words. Similarly to (\ref{kr}),  
 we call it random with respect to $f$ if there is a constant $C_f$ such that
\begin{equation}\label{fr}
n - |f(x_1...x_n)|   < C_f 
 \end{equation}
for all $n$.
We call this deference  the  statistic corresponding to $f$ and define
\begin{equation}\label{fs}
\tau_f(x_1...x_n) = n - |f(x_1...x_n)|   .
 \end{equation}
Similarly,  the sequence $x$ is non-random with respect to $f$ if
\begin{equation}\label{fnon} 
\forall 	C  > 0\, \, \,  \exists \,  n_C  \,\,\, n_C - |f(x_1...x_{n_C} |  \ge C   \, .
\end{equation}
%
%
%
Informally, $x$ is random with respect to $f$ 
 if the statistic $\tau_f$ is bounded by some constant on all prefixes $x_1...x_n$ and, conversely, $x$ is non-random if $\tau_f$ is unbounded when the prefix length grows. 

Based on these definitions, we can reformulate the concepts of randomness and non-randomness in a manner 
 similar to what is customary in mathematical statistics. 
Namely, for any $\alpha\in (0,1)$ we define the set $\{y = y_1...y_n :
\tau_f(y) \ge - \log \alpha \}$. It is easy to see that (\ref{alphasize}) is valid and, therefore, this set represents the critical region $\bar{C}_T(\alpha,n)$, where the test $T$ is as follows: $T = \{ x_1 ... x_n $: 
  $ \tau_f(x_1...x_n) < \alpha \}$.

Based on these consideration,  (\ref{fs}) and the definitions of randomness (\ref{kr}), (\ref{fr}) we   give the following definition of randomness and non-randomness for the  statistic $\tau_f$ and corresponding test $T_f$. 
An infinite sequence $x=x_1x_2 ... $ is random according to the test $T_f$ if there exists such $\alpha > 0$ that for any integer $n$ and this $\alpha$ the word $x_1 ... x_n$ is random (according to the $T_f$ 
 test).  
Otherwise, the sequence $x$ is non-random.

It is important to note that  we can use the statistic
$$ \tau_f =   n- |f(x_1 ... x_n)| $$
with the critical value 
$   t_\alpha = n- \log( 1/\alpha) -1$, $\alpha \in (0,1)$,  
see \cite{rm,rfc}.
So there is no need to use the density distribution formula and it greatly simplifies the use of the test and makes it possible to use this test  for any data compressor $f$.

\subsubsection*{Test performance comparison}

For test $T$, let us define the set $R_T$ of all infinite sequences that are 
  random for  $T$.

We use this definition to compare the ``effectiveness''  of different tests as follows.
The  test $T_1$ is   more efficient than $T_2$ if the size of the difference $R_{T_2} \setminus
  R_{T_1}$ is not equal to zero,  where the size is measured by the Hausdorff dimension. 

Informally, the ``smallest'' set of random sequences corresponds to a test based on Kolmogorov complexity (\ref{kr}) (corresponding set $R_K$ contains ``truly'' random sequences). For a given  test $T_1$ we cannot calculate 
 the difference $R_{T_1} \setminus  R_K$ because the statistic (\ref{kr})  is   noncomputabele, but in the case
of  two tests
 $T_1$ and $T_2$,  where $\dim (R_{T_2} \setminus R_{T_1} )> 0$, we can say that the set of sequences random according to $T_2$ contains clearly non-random sequences. So, in some sense, $T_1$ is more efficient than $T_2$. (Recall that we only consider computable tests.)

The definition of the Hausdorff dimension is given in the Appendix,  but here we briefly note that we use the Hausdorff dimension
 for it  
   as follows: for any binary sequence $x_1x_2 ... $ we define a real number $\sigma(x) = 0.x_1x_2 ... $ and for any set of infinite binary   sequences $S$ we denote the Hausdorff dimension of $\sigma(S)$ by $\dim S$. 
    So, a test $T_1$ is more efficient than $T_2$ (formally $T_1 \succeq T_2$) if $\dim (R_{T_2} \setminus
R_{T_1} ) > 0$.   
Obviously, information about test's effectiveness can be useful to  developers of the tests' batteries.

     Also note that the Hausdorff dimension is widely used in information theory. Perhaps the first such  use was due to Eggleston  \cite{egg} (see also \cite{bi1,bi2}), and later the Hausdorff dimension found numerous applications in AIT \cite{rei,to}.

\subsubsection*{Shannon entropy
}
In RNG testing, one of the popular alternative hypotheses ($H_1$) is that the considered   sequence generated  by Markov process  of memory (or connectivity) 
 $m, m> 0,$ $(S_m)$,  
 but the transition probabilities are unknown. ($S_0$, i.e., $m= 0$,  corresponds to the Bernoulli process). Another popular and perhaps the most general $H_1$ is that the sequence is generated by a stationary ergodic process ($S_\infty$) (excluding $H_0$).

Let us consider the Bernoulli process $\mu \in S_0$ for which $\mu(0) = p, \mu(1) = q,  (p+q=1).$
By definition, the Shannon entropy $h(\mu)$ of  this process 
is defined as 
 $h(\mu) = - (p \log p + q \log q)$ \cite{co}. 
For any stationary ergodic process $\nu \in S$ the entropy of order $k$ is defined as follows:
\begin{multline*}
h_k(\nu) = \\
E_\nu(-\sum_{u \in \{0,1\}^k} (\nu (0/u) \log (0/u) + \nu(1/u) \log \nu(1/u) ) ),
\end{multline*}
where $E_\nu$ is the mathematical expectation according to $\nu$, $\nu(z/u)$ is the conditional probability $\nu (x_{i+1} = z |x_{i-k}... x_i = u) $, 
it does not depend on $i$ due to stationarity \cite{co}.

It is known in Information Theory that for stationary ergodic processes (including $S_\infty$ and $S_m, m \ge 0)$
$h_k \ge h_{k+1}$ for $k\ge0 $ and   there exists  the limit Shannon entropy
$h_\infty (\nu) = \lim h_k(\nu)$.  Besides, for $\nu \in S_m$  $\,\,\, h_\infty = h_m$  \cite{co}.

Shannon entropy plays an important role in data compression because for any lossless and unprefixed code, the average codeword length (per letter) is at least as large  as the entropy, and this limit can be reached.  More precisely, 
let $\phi$ be a lossless,  unprefixed code defined on $\{0,1\}^n, n>0$, and let $\nu \in S$. Then for any $\phi$, $\nu$ and codewords of average length 
\begin{equation}\label{al}
E_n(\phi,\nu)= \frac{1} {n} \sum_{u \in \{0,1\} ^n } \nu(u) |\phi(u)|
\end{equation} 
$E_n(\phi,\nu) \ge h(\nu)$. In addition, there are codes $\phi_1, \phi_2, ... $ such that $\lim_{n\to \infty } E_n(\phi_n,\nu) = 0$ 
\cite{co}. 

\subsubsection*{ Typical sequences and universal codes}
The sequence $x_1x_2...$ is typical for 
 the measure $\mu\in S_\infty$ if for any word $y_1...y_r$ $\lim_{t \to \infty} N_{x_1...x_t}(y_1...y_r)/t = \mu(u)$,
 where    $N_{x_1...x_t }(y_1...y_r) $ is the number of occurrences of a word $y_1...y_r$ in a word  $x_1...x_t $. 

Let us denote the set of all typical sequences as $\mathbf{T}_\mu$ and note that $\mu(\mathbf{T}_\mu) = 1$ \cite{co}.  This notion  is deeply related to information theory. Thus, Eggleston proved the equality $\dim \mathbf{T}_\mu = h(\mu)$ for Bernoulli processes ($\mu \in S_0$) \cite{egg}, and later this was generalized for $\mu \in S_\infty$ \cite{ bi2,rei}.

By definition, a code $\phi$ is universal for a set of processes $S$ if for any $\mu \in S$  and any $x \in \mathbf{T}_\mu $
\begin{equation}\label{un} 
\lim_{n \to \infty} | \phi(x_1 ... x_n)| /n = h_\infty(\mu) .
\end{equation} 

In 1968, R. Krichevsky \cite{krichevsky1968relation} proposed a  code $ \kappa^t_m(x_1 ... x_t)$, $m\ge0, t$ is an integer, whose redundancy, i.e. the average difference between the code length and Shannon entropy, is asymptotically minimal. This code and its generalisations are described in the appendix, but here we note the  following main property.
For any stationary ergodic process $\mu$, that is, $\mu \in S_\infty$ and typical $x \in \mathbf{T}_{\mu}$, 
\begin{equation}\label{ulim}  
\lim_{t \to \infty} |\kappa^t_m(x_1 ... x_t)|/t = h_m(\mu) \, ,
\end{equation}
see  \cite{Krichevsky:93}.

Currently  there are many universal codes which 
are based on different ideas and approaches, among which we note the PPM universal code \cite{cleary1984data},  the arithmetic code \cite{rissanen1979arithmetic},  the Burrows-Wheeler transform \cite{burrows1994block} which is used along with the book-stack (or MTF) code \cite{ryabko1980data,bentley1986locally,ryabko1987technical},  
and some others \cite{sp,re,re2}. 

The most interesting for us is the  class of grammar-based codes suggested by Kieffer and   Yang \cite{kieffer2000grammar,yang2000efficient} which includes the Lempel-Ziv (LZ)  codes
\cite{ziv1977universal} (note that perhaps  the first grammar-based code was described in \cite{kr}). 

The point is that all of  them are universal codes and hence  they ``compress'' stationary processes asymptotically to entropy and therefore 
 cannot be distinguishable at $S_\infty$.  
   On the other hand, we show that grammar-based codes can distinguish "large" sets of sequences as non-random beyond $S_\infty$.

\subsubsection*{Two-faced processes}

The so-called two-faced processes are described in \cite{rm,rfc} and their definitions will be given in Appendix. Here we note some of their properties: the set of two-faced processes $\Lambda_s(p)$ of order $s,  s \ge 1$, and probability $p, p \in (0,1)$, contains the measures  $\lambda$ from $S_s$ such that
$$   
h_0(\lambda) = h_1(\lambda) = ... = h_{s-1}(\lambda) = 1, 
$$ 
\begin{equation}\label{tft}
h_s(\lambda) = h_\infty(\lambda) = -(p\log p + (1-p) \log (1-p) ).
\end{equation} 
Note that they are called two-faced because they appear to be truly random if we look at word frequencies whose length is less than $s$, but are "completely" non-random if the word length is equal to or greater than $s$ (and $p$ is far from $1/2$).

\section{ Comparison of the efficiency of tests for  Markov processes with different memories and general stationary processes.} 

 We now describe the statistical tests  for Markov processes and stationary ergodic processes as follows.
  By (\ref{fs}) statistics definitions are as follows 
$$ \tau_{K^t_m} (x_1 ... x_n) = n -| \hat{\kappa}^t_m(x_1 ... x_n)|,
$$ $$
\tau_{R^t} (x_1 ... x_n) = n -|\hat{\rho}^t(x_1 ... x_n)|
$$
where $\hat{\kappa}^t_m$ and   $\hat{\rho}^t$  are universal codes for $S_m$ and $S_\infty$  defined in  the appendix, see (\ref{unseq}) and (\ref{unseqro}).
We also denote
 the   corresponding tests by $T^t_{K_m}$ and $T^t_{R}$.
The following statement compares the performance of these tests. 

{ \bf Theorem 1.}
For any integers $m,s$ and $t = m s $
$$ T^t_{K_m} \preceq T^t_{K_{m+1}} , \,\, T^t_{K_m}   \preceq T^t_{K_R } .$$
Moreover,  
$\dim(T^t_{K_{m}} \setminus T^t_{K_{m+1}})= 1 $.

{\it Proof.}   
First, let us say a few words about the scheme of the proof. 
 If we apply the $T^t_{K_m}$ test to typical sequences  of a two-faced process $\lambda \in \mathbf{T}_{\Lambda_{{m+1}(p)}}, p\neq 1/2$, they will appear random since 
$h_m(\lambda) =1$, but they are not random according to $T^t_{K_{m+1}}$  since $h_{m+1} (\lambda)$ $=  -(p\log p + (1-p) \log (1-p) ) < 1$ (\ref{tft}). 
So the difference $\dim(T^t_{K_{m}} \setminus T^t_{K_{m+1}}) $ is $-(p\log p + (1-p) \log (1-p) )$
and $\sup_{p \in (0,1/2)} \dim(T^t_{K_{m}} \setminus T^t_{K_{m+1}}) =1$.

More formally, consider a typical sequence $x$ of $\mathbf{T}_{\Lambda_{{m+1}(p)}} $, $p \neq 1/2$.  So,   $\lim_{t \to \infty }-\sum_{u \in \{0,1\}^{m+1}} $ $
(N_{x_1...x_t}(u)/t)  \log(N_{x_1...x_t}(u)/t) = h_\lambda (m) = 1 ,$ see (\ref{tft}), where the first equality is due to typicality, and the second  to the property
 of two-faced  processes  (\ref{tft}).
 
From  here and (\ref{ineq}),  (\ref{unseq}) we obtain $E_\lambda (1/n) | \hat{\kappa}^t_m (x_1 ... x_n)|$ $=1+\epsilon$,
where $\epsilon >0$. From this and typicality we can see that $\lim_{n \to \infty} | \hat{\kappa}^t_m (x_1 ... x_n)|/n$
$= 1+\epsilon$. Hence, there exists such $n_\delta$ that $1+ \epsilon - \delta < | \hat{\kappa}^t_m (x_1 ... x_n)|/n < 1+ \epsilon  +\delta $, if $n > n_\delta$.   So $n -  | \hat{\kappa}^t_m (x_1 ... x_n)| $ $\le n - (n+\epsilon - \delta)$.
So, if we take $\delta = \epsilon /2$, we can see that for $n>n_\delta$  $n -  | \hat{\kappa}^t_m (x_1 ... x_n)| $ is negative. From this and the definition of randomness (\ref{fr}) we can see that 
typical sequences from $\mathbf{T}_{\Lambda_{{m+1}(p)}} $ are random according to 
$ \hat{\kappa}^t_m (x_1 ... x_n)$, i.e.  $ T^t_{K_m}$. 
From this and (\ref{kr-le}) we obtain  $T^t_{K_{m+1}} \preceq T^t_{R }$.

\section{Effectiveness of tests based on Lempel-Ziv data compressors} 
In this part we   will describe a test that is more effective than $T_{R}^t$ and $ T^t_{K_m}$ for any $m$.

First, we will briefly describe the LZ77 code based on the definition in \cite{gram}.
Suppose, there is a binary string $\sigma^*$ which is encoded using the code LZ77. 
This string  is represented by a list of pairs
$(p_1; l_1) ...  (p_s; l_s)$. Each pair $(p_i; l_i)$ represents a string,
and the concatenation of these strings is $\sigma^*$. In particular,
if $p_i = 0$, then the pair represents the string $l_i$, which is
a single terminal. If $p_i  \neq 0$, then the pair represents a
portion of the prefix of $\sigma^*$ that is represented by the preceding
$i - 1$ pairs; namely, the $l_i$ terminals beginning at
position $p_i$ in $\sigma^*$; see \cite[part 3.1]{gram}.
The length of the codeword depends on the encoding of the sub-words $p_i, l_i$
which are integers. For this purpose we will use a prefix code $C$ for integers, for which for any
integer $m$  
\begin{equation}\label{C}
   |C(m)| = \log m +2 \log \log (m +1) +O(1)  \,\,.
 \end{equation}
 Such  codes are known in Information Theory, see, for example,  \cite[part 7.2]{co}. 
 Note that $C$ is the prefix code and, hence,  for any $r \ge 1$ the codeword $C(p_1)C(l_1) ... C(p_r) C(l_r)$ can be decoded to $(p_1;l_1) ... (p_r; l_r)$.
There  is the following upper bound for the length of the LZ77 code \cite{co,gram}: for any
word $w_1 w_2 .... w_m$  
\begin{equation}\label{leng2}
   | code_{LZ}(w_1 w_2 ... w_m)| \le m \, (1+o(1) ),
 \end{equation}
 if $m \to \infty$.

We will now describe such   sequences that, on the one hand,   are not  typical for any stationary ergodic measure and, on the other hand,  are not   random and will be rejected by the suggested test. Thus, the proposed model allows us to detect non-random sequences that are not typical for for any stationary processes.
 On the other hand, those sequences  are recognized tests based on  LZ77  as non-random.
To do this, we take any  random sequence $ x = x_1 x_2 ... $  
(that is, for which (\ref{kr}) is valid) 
and define a new sequence $ y (x) = y_1 y_2. .. $ as follows. Let for   $k =0,1, 2, ... $ 
 \begin{equation}\label{Y}
  u_k = x_{2^{2^k}-1} x_{2^{2^k}} x_{2^{2^k}+1 }... x_{2^{2^{k+1}}-2}
$$ $$
y(x)= u_0 u_0 u_1 u_1  u_2  u_2 u_3 u_3  ...
\end{equation}
(For example, $u_0 = x_1 x_2 $, $u_1 = x_3$ $ x_4$ $ ... $ $x_{14}$, $u_2 = x_{15} ... x_{254}$, 
$y(x)=  x_1 x_2  $ $x_1 x_2   x_3 x_4 ... x_{14} $ $ x_3 x_4 ... x_{14}$ 
$x_{15} ... x_{254}$ $x_{15} ... x_{254}$ $...$. 

The idea behind this sequence is quite clear. Firstly, it is obvious that the word $y$ 
cannot be typical for a stationary ergodic source and,  
secondly, when $u_0 u_0 u_1 u_1 ...  u_k u_k$ is encoded 
 the second subword $ u_k$ will be encoded by a very short word (about $ O (\log |u_k|))$, since it coincides with the previous word  $u_k$. So, for large $k$ the length of 
 the encoded word $LZ(u_0 u_0 u_1 u_1  ...  u_k u_k)$ will be about 
 $|u_0 u_0 u_1 u_1 ...  u_{k} u_k| \, (1/2+ o(1) )\, $.
So $\lim \inf_{n \to \infty} | LZ  (y_1 y_2 ... y_n) |  /n = 1/2$.
Hence it follows that
 \begin{equation}\label{1/2}
\dim(\{y(x): x \, is \, random \} ) = \, 1/2.
 \end{equation} 
(Here we took into account that $x$ is random and,   $\dim \{x: x \,\, is \,\, random \} =1$, see \cite{rei}).
So,   having taken into account the definitions of non-randomness  (\ref{fs}) and (\ref{fnon}) we can see that    $y(x)$ is non random according to  statistics $\tau = n-| LZ(y_1 ... y_n)| $. Denote this test by $T_{LZ}$. 

Let us consider the test $T^t_{K_m}$, $m,t$ are integers.  Having taken into account that the sequence $x$ is random, we can see   that $\lim_{t \to \infty} $ $|\kappa_m^t(x_i x_{i+1} ... x_{i+t}| / t =1.$ 
So, from from   (\ref{unseq}) we can see that for any $n$
$| \hat{\kappa}_m^t(x_1 ... x_{n}| / t  =1 +o(1)$.
The same reasoning is true for the code $\hat{\rho}^t$. 

We can now compare the size of random sequence sets across different tests as follows: 
$$ R_{T^t_{K_m}} \setminus  R_{T_{LZ}} \supset \{y(x): x \, is \, random \} \, .
$$
Taking into account  (\ref{1/2}) we can see that
$$ \dim( R_{T^t_{K_m}} \setminus  R_{T_{LZ}} ) \ge 1/2 \, .  $$
Likewise, the same is true for the $T_R$ test.
From the latest inequality  we obtain the following 

{\bf Theorem 2
.}  
For any random (according to (\ref{kr}) ) 
sequence $x$  the sequence $y(x)$ is non-random for  
the test $ T_{LZ}$, whereas this sequence is random for tests $T^t_{R}$   and $T^t_{K_m}$. Moreover,  
  $T^t_{R } \preceq T_{LZ }$ and $T^t_{K_m}  \preceq T_{LZ }$ for any $m,t$.

{\bf Comment.}  The sequence $y(x)$ is constructed by duplicating parts of $x$. This construction can be slightly modified as follows: instead of duplication (as $u_i u_i$), we can use $u_i u_i^\gamma $, where $ u_i^\gamma$ contains $\gamma |u|$ the first letters of $u$, $\gamma <1/2$. In this case
$ \dim( R_{T^t_{K_m}} \setminus  R_{T_{LZ}} ) \ge1-\gamma \, $ and, therefore, 
$$ \sup_{\gamma \in (0,1/2) } \dim( R_{T^t_{K_m}} \setminus  R_{T_{LZ}} ) = 1\, .$$

\section{Conclusion}
Here we describe some recommendations for practical testing of RNG. 
 Based on Theorem 1, we can recommend to use several tests $T_{K_s^t}$,  
  based on the analysis of occurrence frequencies of words of different length $s$. 
  In addition, we recommend using tests for which $s$ depends on the length $n$  
   of the sequence under consideration. For example, $s_1 = O(\log \log n)$), $s_2 = O(\sqrt{\log n}$), etc. They can be included in the test battery directly or as  the ``mixture'' $T_R$ 
with several non-zero $\beta$ coefficients, see (\ref{beta}) in the Appendix.
 
Theorem 2 shows that it is useful to include tests based on dictionary data compressors such as the Lempel-Ziv code.     In such a case we can use the statistic
$$ \tau_{LZ} =   n- | LZ(y_1 ... y_n)| $$
with the critical value 
$   t_\alpha = n- \log( 1/\alpha) -1$, $\alpha \in (0,1)$,  
see \cite{rm,rfc}.
Note that in this case there is no need to use the density distribution formula, which greatly simplifies the use of the test and makes it possible to use a similar test for any grammar-based data compressor.


\section{Appendix}
\subsection{Hausdorff dimension}
Let  $A \subset [0,1], \rho >0$.
A family of sets $S$  is called a $\rho$-covering $A$ if

i) $S$ is finite or countable, ii) any $\sigma \subset [0,1] $ and its  length is not grater than $\rho$ and  iii)
$\cup_{\sigma \in S} \sigma  \supset A$.
Let 
$$ l(\alpha,A, \rho) = \inf \sum_{\sigma \in S}  diam(\sigma)^\alpha, $$
where the infimum is taken over all $\rho$-coverings. Then Hausdorff 
 dimension $\dim(A)$  is determined by the equality
$$ \dim(a) = \inf_{\alpha}  \lim_{\rho \to 0} l(\alpha,A, \rho) =0 = \sup_{\alpha}  \lim_{\rho \to 0} l(\alpha,A, \rho) = \infty  .$$

\subsection{ Krichevsky universal code and twice-universal code.}

Krychevsky in \cite{krichevsky1968relation} described the following measure $K_0$ and universal code $\kappa_0$ for Bernoulli processes, which in the case of the binary alphabet looks like this
$$
K_0^t (x_1 x_2 ... x_t ) =   \prod_{i=0}^{t-1}   \frac{ N_{x_1...x_i }(x_{i+1})  +1/2 )} {i+1} ,  
$$ $$\kappa_0^t (x_1 x_2 ... x_t ) = \lceil -\log K_0(x_1 x_2 ... x_t ) \rceil \, .
$$
Then he generalised them for Markov chains of memory $m, $ $m>0$ \cite{Krichevsky:93}, as follows
\[ K_m^t (x_1  ... x_t ) = 
 \begin{cases}
     \frac{1}  {2^{t}}   &  \\  \qquad  \qquad   \qquad  \qquad    \text{if } t\le m\\ 
     \frac{1}  {2^{t}}  \prod_{i=m}^{t-1} \frac{N_{x_1... x_i}(x_{i+1-m} ... x_{i+1} )+1/2} 
{N_{x_1... x_{i-1}}(x_{i+1-m} ... x_{i} )+1}
 &\\  \qquad  \qquad   \qquad  \qquad  \text{if } t>m, \end{cases}
\]
$ \kappa_m^t (x_1  ... x_t ) = \lceil   - \log K_m^t (x_1  ... x_t ) \rceil $, see \cite{Krichevsky:93}.
For example,   $$K^5_0(01010) = \frac{1/2}{1}\frac{1/2}{2}\frac{132}{3}\frac{3/2}{4} \frac{5/2}{5}, 
$$ $$
K^5_1(01010) =
\frac{1} {2} \frac{1/2} {1} \frac{1/2} {1} \frac{3/2} {2} \frac{3/2} {2} \, .$$

The code $\kappa_m^t$ is universal for a set of processes $S_m$, and , for any $\nu \in S_m$  
\begin{equation}\label{ineq} 
   h_m (\nu)    <  E_\nu(\kappa^t_m,\nu) \le h_m (\nu) + 2^m \log t /(2t) +O(1/t)
\end{equation} \cite{krichevsky1968relation,Krichevsky:93}. 
(This code is optimal in the sense that the  redundancy,  that is   $2^m \log t /(2t) +(1/t)$,
 is asymptotically minimal \cite{krichevsky1968relation,Krichevsky:93}).

One of the first universal codes for the set of all stationary ergodic processes $S_\infty$ was proposed in \cite{tu}. For this code, the measure $\rho$ and the code length $R$ are defined as follows:
\begin{equation}\label{beta} 
R^t(x_1...x_t) = \sum_{i=0}^\infty  \beta_{i} K_i^t (x_1 x_2 ... x_t ), 
\end{equation} 
$$ \rho^t(x_1...x_t) = \lceil -\log R^t(x_1 x_2 ... x_t ) \rceil \, ,
$$
where $\sum_{i=0}^\infty \beta_i = 1$ and $\forall i : \beta_i >0$.
Obviously,  for any $j$ 
$$ -\log \sum_{i=0}^\infty  \beta_{i} K_i^t (x_1 x_2 ... x_t ) 
=  - \log  \beta_{j} K_j^t (x_1 x_2 ... x_t ) + $$ $$   -\log  (1+ \sum_{i=0,  i \neq j}^\infty  \beta_{i} K_i^t (x_1 x_2 ... x_t ) / (\beta_{j} K_j^t (x_1 x_2 ... x_t  ) ) 
$$ $$
 \le - \log 
 \beta_{j} K_j^t (x_1 x_2 ... x_t )  \, .$$
Hence, 
$$\rho^t(x_1...x_t) \le  \lceil - \log \beta_{j}    -\log K_j^t (x_1 x_2 ... x_t )  \rceil \le  \lceil - \log \beta_{j}   \rceil$$ 
\begin{equation}\label{kr1} 
 + \lceil -\log K_j^t (x_1 x_2 ... x_t )  \rceil =  \lceil - \log \beta_{j}   \rceil + | \kappa_j^t(x_1 x_2 ... x_t )| .
\end{equation}
This code is called twice universal \cite{tu} because it can be used to compress data when both the process memory and the probability distribution are unknown.

Usually, when using universal codes, the sequence $x_1...x_n$ is encoded in parts as follows:
\begin{equation}\label{unseq} 
\hat{\kappa}^t_m(x_1 ... x_n) = \kappa^t_m(x_1 ... x_t)\kappa^t_m(x_{t+1} ... x_{2t}).... \kappa^t_m(x_{n-t+1}... x_n) 
\end{equation}
(for brevity, we assume that $n/t$ is an integer). Let us similarly define
\begin{equation}\label{unseqro} 
\hat{\rho}^t        
(x_1 ... x_n) = \rho^t(x_1 ... x_t)\rho^t(x_{t+1} ... x_{2t}).... \rho^t(x_{n-t+1}... x_n) 
\end{equation}
Taking into account the definition of $\kappa_j^t (x_1 x_2 ... x_t )$ and equations (\ref{kr}),  (\ref{unseq}), (\ref{unseqro}) we obtan that for any integer $j$
\begin{equation}\label{kr-le} 
\hat{\rho}^t  (x_1 ... x_n) \le \hat{\kappa}^t_j(x_1 ... x_n)  +O(n/t).
\end{equation}

\subsection{ Two-faced processes}

Let us first we consider several examples  of  two-faced  Markov chains. 
Let a matrix of transition probabilities $T_1 $ be as follows:
$$
T_1 = \qquad
\begin{tabular}{c| c c }
$\qquad  $& 0  & 1   \\
\hline 0 &
 $\nu$ & $1-\nu$ 
\\
 1&   $1-\nu$ & 
$\nu$ \\
\end{tabular} 
\, \,  \, \,  \, \,  \, \,  \, \,  \, \,   , 
$$ $$
$$
where $\nu \in (0,1)$ 
(i.e.  
$P\{ x_{i+1} = 0| x_i = 0 \} = \nu$, $P\{ x_{i+1} = 0| x_i = 1 \} = 1-\nu , ... $). 
The "typical" sequences for  $\nu= 0.9$   and $\nu= 0.1 $  can be as follows:
$$ 0000000000 \, \,  \,111111111 \, \,  \,0000000000 \, \,  \,1111111 \, \,  \,0 \, ...  \, \, ,
$$
$$  01010101 \, \,\, 1010101010 \,\,\, 010101010101010101 \,\,\,1010 \, ... \, \, \,  .$$
(Here the gaps correspond to state 
transitions.) Of course, 
these sequences  are not truly random. On the other hand, the frequencies of 1's and 0's go to $1/2$ due to the symmetry of the matrix  $T_1$.  

Define
$$
\hat{T}_1 = \qquad
\begin{tabular}{c| c c }
$\qquad  $& 0  & 1   \\
\hline 0 &
 $1-\nu$ & $\nu$ 
\\
 1&   $\nu$ & 
$1-\nu$ \\
\end{tabular} 
$$ $$
$$
$$T_2 =  (T_1 \hat{T}_1) =
\begin{tabular}{c| c c c c}
$\qquad  $& 00  & 01 & 10 & 11  \\
\hline 0 &
 $ \nu$ & $1- \nu$ & $1 - \nu$ & $\nu$ 
\\
 1&   $1-\nu$ & 
$\nu$ & $ \nu$ & $1-\nu$ \\
\end{tabular}
$$
\\ \\
(Here   $ \, \, P\{ x_{i+1} = 0| x_i = 0, x_{i-1}=0 \} = \nu$, $P\{ x_{i+1} = 0| x_i = 0, x_{i-1}=1 \} = 1-\nu , ... \, $.).

 
Now we can define 
  a transition matrix two-faced Markov chains with differen memory  as follows.

 The  $k+1$-order transition matrix $T_{k+1} = T_k \hat{T_k}$,  $\hat{T}_{k+1} = \hat{T_k} T_k $, $k=2, 3, ...$. 
T In order to define  the process  $x_1 x_2 ...$ the initial probability distribution needs to be specified. 
We define 
the initial distribution of the processes $T_k$ and $\bar{T}_k$,  $k=1,2, \ldots\,$,
to be uniform on $\{0,1\}^k$, i.e. $P\{ x_1 ... x_k = u \} = 2^{-k}$ for any $u \in \{0,1\}^k$. 

 The following  statement from \cite{rm,rfc} describes the main properties of 
the   processes defined above.\\ 
{\bf Claim.}\label{T1}
Let a sequence $x_1 x_2 ... $ be 
generated by the process  $T_k$ (or $\bar{T}_k$), $k \ge 1$
and $u$ be a binary word of length $k$.
Then,  if the initial state obeys the uniform distribution over $\{0,1\}^k$, then 
\begin{itemize}
 \item[i)]
for any $j \ge 0$ 
 \begin{equation}\label{u2}  
 P(x_{j+1} ... x_{j+k} = u) = 2^{-|u|}  .
 \end{equation}  
\item[ii)]
 For each $\nu \in (0,1) $ the $k$-order
Shannon entropy ($h_k$) of the processes $T_k$ and
$\bar{T}_k,$ equals 1 bit per letter 
whereas the limit Shannon entropy ($h_\infty $) equals $ - (\nu
\log_2 \nu + (1- \nu) \log_2 (1-\nu) ).$ 
\end{itemize}



\end{document}